# The Restriction Condition of Gauge Transformation Which the Motion Equation of Non-Abelian Gauge Field Must Satisfy and Elimination of Higgs Mechanism

Mei   Xiaochun

(Institute of Innovated Physics in Fuzhou, China, E-mial:mxc001@163.com)

**Abstracts** In the current theory of non-Abelian gauge field, we only claim the invariability of Lagrangian, without claim the invariability of the motion equation. This is inconsistent and irrational. It is proved that a restriction relation between gauge potentials and group parameters must be satisfied in order to ensure the gauge invariability of the motion equation of non-Abelian gauge field, and the restriction relation is equivalent to the Faddeev--Popov theory. The result leads to that the completely local gauge invariability is violated but there still exists the incompletely local gauge invariability. After the restriction relation is considered, the mass item of the non-Abelian gauge fields can be added into the Lagrangian and motion equation directly without violating gauge invariability. The corresponding $W, T$ identity is obtained and the theory is still renormalizable. In this way, the Higgs mechanism becomes unnecessary. It means that we can reach a coincident theory without the hypothesis of the Higgs particles again. The description of the stander theory of particle physics can also be simplified greatly and the problem of $CP$ violation in strong interaction can also be solved thoroughly.

## 1. The Restriction Condition of Gauge Transformation Which Motion Equation of Non-Abelian Gauge Field Must Satisfy

According to the Yang—Mills theory, in order to keep the Lagrangian unchanged under the local gauge transformation, the transformation rules of the field $\phi$ and its covariant differentiation should be defined as

$$\phi'(x) = exp\{-i\theta^\alpha(x)T^\alpha\}\phi(x) \tag{1}$$

$$D'_\mu(x)\phi'(x) = exp\{-i\theta^\alpha(x)T^\alpha\} D_\mu(x)\phi(x) \tag{2}$$

$$D_\mu(x) = \partial_\mu + A_\mu(x) \qquad A_\mu(x) = -igA_\mu^\alpha(x)T^\alpha \tag{3}$$

Here $\theta^\alpha(x)$ are group parameters. The function forms of $\theta^\alpha(x)$ are considered arbitrary at present. From Eq.(2), we can get the transformation rule of gauge potential $A_\mu^\alpha$

$$A_\mu^{\prime\alpha}(x) = A_\mu^\alpha(x) + f^{\alpha\beta\gamma}\theta^\beta(x)A_\mu^\gamma(x) - \frac{1}{g}\partial_\mu\theta^\alpha(x) \tag{4}$$

The intensities of gauge fields are defined as

$$F^\alpha_{\mu\nu}(x) = \partial_\mu A^\alpha_\nu(x) - \partial_\nu A^\alpha_\mu(x) + g f^{\alpha\beta\gamma} A^\beta_\mu(x) A^\gamma_\nu(x) \tag{5}$$

Its transformation rule is

$$F'^\alpha_{\mu\nu}(x) = F^\alpha_{\mu\nu}(x) + f^{\alpha\beta\gamma} \theta^\beta(x) F^\gamma_{\mu\nu}(x) \tag{6}$$

The Largrangian of non-Abelian gauge fields with zero masses

$$\mathcal{L}_0(x) = -\frac{1}{4} F^\alpha_{\mu\nu}(x) F^\alpha_{\mu\nu}(x) \tag{7}$$

is unchanged under the gauge transformation. But the Lagrangian with mass items can't keep unchanged under the gauge transformation.

On the other hand, there is no discussion about whether or not the motion equation of gauge field with zero mass is unchanged under the gauge transformation (4) at present, or this problem is actually neglected. Physicists seem to think that it is enough that the Largrangian of gauge field is unchanged under the gauge transformation. However, it should be pointed out that the invariability of the Largrangian can not ensure the invariability of the motion equation of gauge field, because both are independent each other. In order to deduce the motion equation from the Largrangian, we need to use the Largrangian equation. But the Largrangian and Largrangian equation are two different things. The concrete forms of gauge fields are actually determined by the motion equation. We can not determine the concrete forms of gauge fields only based on the Largrangian. The Largrangian is only a part of the motion equation corresponding to the interaction energy. In principle, the motion equation is more essential than the Largrangian. If we want to keep the gauge theory unchanged under the gauge transformation, we should make both the Largrangian and the motion equation unchanged simultaneously. Otherwise the theory can not be considered consistent only to consider the invariability of the Largrangian without considering the invariability of the motion equation. As shown below that as soon as the invariability of the motion equation is taken into account, a great influence on the gauge theory would be caused. At present, we always think that the forms of group parameters $\theta^\alpha(x)$ can be arbitrary in the local gauge transformation of non-Abelian gauge field. It will be proved below that after the gauge transformations of non-Abelian gauge field's motion equation are carried out, a certain restriction conditions would be introduced consequentially, indicating that completely local gauge invariability should be changed into the incompletely local gauge invariability. The result is that we can describe the gauge particles with non-zero mass consistently without using the Higgs mechanism. The description of the stander theory of particle physics can also be simplified greatly and the problem of $CP$ violation in strong interaction can also be solved thoroughly.

Let's first take free electromagnetic field as an example to show the existence of constriction condition. By considering the Lorentz condition $\partial_\mu A_\mu = 0$, the motion equation of free electromagnetic field is

$$\partial_\mu F_{\mu\nu} = \partial^2 A_\nu - \partial_\nu \partial_\mu A_\mu = \partial^2 A_\nu = 0 \tag{8}$$

The $U(1)$ gauge transformation is defined as

$$A'_\nu(x) = A_\nu(x) - \frac{1}{g} \partial_\nu \theta(x) \tag{9}$$

So the gauge transformation of Eq. (8) is

$$\partial_\mu F'_{\mu\nu} = \partial^2 A'_\nu(x) = \partial^2 A_\nu(x) - \frac{1}{g} \partial_\nu \partial^2 \theta(x) = 0 \tag{10}$$

By considering the motion equation (8) again, we get the restriction condition for $U(1)$ group parameter

$$\partial_\nu \partial^2 \theta(x) = 0 \tag{11}$$

It means that $\theta =$ constant, $\partial_\nu \theta =$ constant, or $\partial^2 \theta =$ constant, (If supposing that the Lorentz condition $\partial_\mu A_\mu = 0$ is unchanged under gauge transformation, we get $\partial^2 \theta = 0$). So the forms of group parameter can't be arbitrary. It should be noted that this constriction condition is introduced automatically when the gauge transformation of motion equation is carried out. The condition is not an artificial hypothesis and has nothing to with gauge potentials.

When there is interaction between electromagnetic field and spinor field, the motion equation of electromagnetic field is

$$\partial^2 A_\mu(x) = -j_\mu \qquad j_\mu = i\frac{e}{2}\left(\overline{\psi}\gamma_\mu \psi - \psi^\tau \gamma_\mu^\tau \overline{\psi}^\tau\right) \tag{12}$$

Because the form of flow $j_\mu$ is unchanged under the transformation shown in Eq.(1), the $U(1)$ gauge transformation of Eq.(12) is

$$\partial^2 A_\mu(x) - \frac{1}{g}\partial_\mu \partial^2 \theta(x) = -j_\mu \tag{13}$$

It is obvious that the restriction condition Eq.(11) is necessary to keep the non-free motion equation of electromagnetic field unchanged under the transformation. That is to say, when we do the gauge transformation of the motion equation of free electromagnetic field, the restriction condition is introduced. It is just this condition that ensures the motion equation of non-free electromagnetic field unchanged under gauge transformation. The result is self-consistent.

For general $U(1)$ gauge fields, there exists no Lorentz condition with $\partial_\mu A_\mu \neq 0$, the $U(1)$ gauge transformation of free field's motion equation is

$$\partial_\mu F'_{\mu\nu} = \partial^2 A'_\nu - \partial_\nu \partial_\mu A'_\mu = \partial^2 A_\nu - \partial_\nu \partial_\mu A_\mu = \partial_\mu F_{\mu\nu} = 0 \tag{14}$$

So the group parameter $\theta(x)$ can take arbitrary form. The non-free motion equation can also keep unchanged under the transformation.

For $SU(N)$ non-Abelian gauge field without considering mass item, the motion equation is[1]

$$\partial_\mu F^\alpha_{\mu\nu} + gf^{\alpha\beta\gamma} A^\beta_\mu F^\gamma_{\mu\nu} = 0 \tag{15}$$

Under the gauge transformation defined in Eqs.(4) and (6), the motion equation becomes

$$\partial_\mu F'^\alpha_{\mu\nu} + gf^{\alpha\beta\gamma} A'^\beta_\mu F'^\gamma_{\mu\nu} = \partial_\mu \left(F^\alpha_{\mu\nu} + f^{\alpha\beta\gamma}\theta^\beta F^\gamma_{\mu\nu}\right)$$

$$+ gf^{\alpha\beta\gamma}\left(A^\beta_\mu + f^{\beta\rho\sigma}\theta^\rho A^\sigma_\mu - \frac{1}{g}\partial_\mu \theta^\beta\right)\left(F^\gamma_{\mu\nu} + f^{\gamma\lambda\omega}\theta^\lambda F^\omega_{\mu\nu}\right) = 0 \tag{16}$$

By considering Eq.(15) again, the formula above becomes

$$f^{\alpha\beta\gamma}\left\{\left(\partial_\mu \theta^\beta\right)F^\gamma_{\mu\nu} + \theta^\beta \partial_\mu F^\gamma_{\mu\nu}\right\} + gf^{\alpha\beta\gamma}f^{\gamma\lambda\omega} A^\beta_\mu \theta^\lambda F^\omega_{\mu\nu}$$

$$+ gf^{\alpha\beta\gamma}\left(f^{\beta\rho\sigma}\theta^\rho A^\sigma_\mu - \frac{1}{g}\partial_\mu \theta^\beta\right)\left(F^\gamma_{\mu\nu} + f^{\gamma\lambda\omega}\theta^\lambda F^\omega_{\mu\nu}\right) = 0 \tag{17}$$

It is easy to see that the solution of formula above is

$$\partial_\mu \theta^\alpha = g f^{\alpha\beta\gamma} \theta^\beta A_\mu^\gamma \tag{18}$$

In fact, by considering Eqs.(15) and (18), as well as the anti-symmetry relation of group construction constant $f^{\alpha\beta\gamma}$, the left side of Eq.(17) becomes

$$f^{\alpha\beta\gamma} f^{\beta\rho\sigma} \theta^\rho A_\mu^\sigma F_{\mu\nu}^\gamma - f^{\alpha\beta\gamma} f^{\gamma\rho\sigma} \theta^\beta A_\mu^\rho F_{\mu\nu}^\sigma + f^{\alpha\beta\gamma} f^{\gamma\rho\sigma} \theta^\rho A_\mu^\beta F_{\mu\nu}^\sigma$$

$$= -\left( f^{\alpha\sigma\gamma} f^{\gamma\rho\beta} + f^{\alpha\rho\gamma} f^{\gamma\beta\sigma} + f^{\alpha\beta\gamma} f^{\gamma\sigma\rho} \right) \theta^\rho A_\mu^\beta F_{\mu\nu}^\sigma \tag{19}$$

By means of the Jacobian relation $f^{\alpha\sigma\gamma} f^{\gamma\rho\beta} + f^{\alpha\rho\gamma} f^{\gamma\beta\sigma} + f^{\alpha\beta\gamma} f^{\gamma\sigma\rho} = 0$, we prove that the Eq.(18) is the solution of Eq.(17). Eq.(18) is just the restriction condition that group parameters of $SU(N)$ non-Abelian gauge field should satisfied. **It should be emphasized again that this constriction condition is introduced naturally when we do the gauge transformation of free non-Abelian gauge field's motion equation. It is not an artificial hypothesis.** Meanwhile, the relation (18) is relative to gauge potentials. The situation is different from the Abelian gauge theory.

Similarly, for non-free $SU(N)$ gauge fields without considering mass item, the motion equation is

$$\partial_\mu F_{\mu\nu}^\alpha + g f^{\alpha\beta\gamma} A_\mu^\beta F_{\mu\nu}^\gamma = -ig \overline{\psi} \frac{\lambda^\rho}{2} \gamma_\mu \delta_{\mu\nu} \delta^{\alpha\rho} \psi \tag{20}$$

It is obvious that only when group parameters satisfy restriction condition (18), the motion equation can keep unchanged under $SU(N)$ transformation. In the current non-Abelian gauge theory, however, we only consider the invariability of the Lagrangian under gauge transformation, without considering the invariability of motion equation. This situation is unacceptable. In fact, if the restriction condition (18) is not considered, after $SU(N)$ transformation, the group parameters $\theta^\alpha(x)$ with arbitrary forms would appear in the motion equation. This kind of motion equation is meaningless in physics. It is just the restriction condition that can ensure both the motion equation and the Lagrangian unchanged under $SU(N)$ transformation simultaneously so that theory becomes self-consistent.

It can be known from Eq.(4) that Eq.(18) just means $A_\mu'^\alpha = A_\mu^\alpha$, i.e., the gauge potential themselves are unchanged under $SU(N)$ transformation. For $SU(N)$ gauge $\alpha = 1,2,\cdots N$, Eq.(18) represents $4N$ equations. Because we only have $N$ group parameters, so the function forms of group parameters are not unique. We can have several selections to satisfy Eq.(18). For example, for $SU(3)$ gauge group, there are 4 differential equations for parameter $\theta^1$. We can let three group parameters satisfy the first three equations so that this set of parameters is determined completely. Then let the second set of group parameters $\theta^1, \theta^2, \theta^3$ satisfies the fourth differential equations. In this set of parameters, two of three are independent. We can also think that each of all four differential equations is independent. Because each equation contains three group parameters, two of them are independent, and so do. Because $\theta^\alpha(x) \neq$ constant in this cases, even though gauge potential themselves are unchanged under the transformation, the gauge transformations of other fields $\phi(x)$ and their covariant differentials defined in Eqs.(1) and (2) are still meaningful. But the group parameter's form of each set would be restricted by Eq.(18) so that thy can not be completely independent and arbitrary again.

We call the gauge theory in which group parameters are arbitrary as the completely local gauge theory, the gauge theory in which group parameters are not completely arbitrary as the incompletely local gauge theory. It can be said in general that the completely local gauge invariability is actually impossible, for it would destroy the invariability of gauge field's motion equations. The arbitrary group parameters would appear in the motion equations after gauge transformation so that the motion equations become meaningless.

It should be emphasized again that the constriction conditions (11) and (18) are not additional hypotheses. They are introduced naturally when the gauge transformations of gauge field's motion equations are carried out. Only based on them, the gauge theory becomes meaningful and self-consistent. It will be proved below that as long as the principle of completely local gauge invariability is replaced by the principle of incompletely local gauge invariability, we can describe the gauge particles with non-zero mass consistently without introducing the Higgs mechanics. That is to say, the Higgs mechanics would become unnecessary. The restriction relation does not cause any inconsistency that contradicts the current experiments, and the description of gauge theory would become more symmetrical and simple.

It is shown below that the result above coincides with the Faddeev—Popov theory [2]. In order to avoid infinity in the theory of gauge fields, Faddeev and Popov suggested that the orbit integral over function space should be restricted on the hyper-surface decided by the gauge condition $F(A_\mu^\alpha) = 0$, $\alpha = 1, 2, \cdots N$. In this way, the freedom degrees of gauge fields are decreased to $3N$ from $4N$. The following relation is used to restrict orbit integral

$$\Delta_F(A_\mu^\alpha) \cdot \int [dg] \, \delta\{F(A_\mu^\alpha)\} = 1 \tag{21}$$

The restriction condition $\delta\{F(A_\mu^\alpha)\}$ demands $F(A_\mu^{\alpha g}) = 0$ actually. For $U(1)$ gauge field, by taking the Landau gauge condition $F(A_\mu^\alpha) = \partial_\mu A_\mu = 0$, we have

$$F(A_\mu^g) = 0 \rightarrow \partial_\mu A_\mu' = \partial_\mu \left( A_\mu - \frac{1}{g} \partial_\mu \theta \right) = -\frac{1}{g} \partial^2 \theta = 0 \tag{22}$$

The result coincides with Eq.(11). For non-Abelian gauge fields, by using the Landau gauge condition $\partial_\mu A_\mu^\alpha = 0$, we have

$$F(A_\mu^{\alpha g}) = 0 \rightarrow \partial_\mu A_\mu'^\alpha = \partial_\mu \left( A_\mu^\alpha + f^{\alpha\beta\gamma} \theta^\beta A_\mu^\gamma - \frac{1}{g} \partial_\mu \theta^\alpha \right)$$

$$= \partial_\mu \left( f^{\alpha\beta\gamma} \theta^\beta A_\mu^\gamma - \frac{1}{g} \partial_\mu \theta^\alpha \right) = 0 \tag{23}$$

The formula can be written as

$$f^{\alpha\beta\gamma} \theta^\beta A_\mu^\gamma - \frac{1}{g} \partial_\mu \theta^\alpha = b_\mu \qquad \partial_\mu b_\mu = 0 \tag{24}$$

Taking the simplest form to let $b_\mu = 0$, we reach Eq.(18). So the result in this paper coincides with the Faddeev—Popov theory, or the constriction conditions introduced in the paper is the simplest form of the Faddeev—Popov theory. It can be seen that though restriction relation (23) can eliminate the infinity of orbit integral, it can't make motion equation invariable under gauge transformation. To make both the orbit integral finite and the motion equation of non-Abelian gauge fields invariable, the restriction relation $\delta(\partial_\mu b_\mu)$ should be changed into $\delta(b_\mu)$ further. So Eq.(21) should be re-written as

$$\Delta_F(A_\mu^\alpha) \cdot \int [dg] \, \delta\left[ R_\mu \left( f^{\alpha\beta\gamma} \theta^\beta A_\mu^\gamma - \frac{1}{g} \partial_\mu \theta^\alpha \right) \right] = 1 \tag{25}$$

Here $R_\mu$ is an arbitrary constant vector to ensure Eq.(18) always tenable. In this way, by the relation $\Delta_F(A_\mu^\alpha) = \det M_F$ we have

$$M_F^{\alpha\beta}(x,y) = R_\mu \left[ f^{\alpha\beta\gamma} A_\mu^\gamma(y) \delta^4(x-y) - \frac{1}{g} \delta_{\alpha\beta} \partial_\mu \delta^4(x-y) \right] \tag{26}$$

The action of ghost particle corresponding to $SU(2)$ gauge fields becomes

$$S_g = \int d^4 x C_\alpha^+ R_\mu \left( \delta_{\alpha\beta} \partial_\mu - gf^{\alpha\beta\gamma} A_\mu^\gamma \right) C_\beta \tag{27}$$

It can be seen below that this kind of change has no essential influence on the theory for ghost particles are fictitious actually. The action of ghost particle corresponding to $U(1)$ gauge fields is unchanged.

Because the non-Abelian gauge potentials themselves should be unchanged with $A'^\alpha_\mu = A^\alpha_\mu$ under $SU(2)$ gauge transformation, so the Lagrangian with mass items is invariable under gauge transformation

$$\mathcal{L} = -\frac{1}{4} F_{\mu\nu}^\alpha F_{\mu\nu}^\alpha - \frac{1}{2} m_A^2 A_\mu^\alpha A_\mu^\alpha \tag{28}$$

That is to say, we can add mass items into the Lagrangian and motion equation directly without violating gauge invariability. It will be proved below that when the interaction between gauge particles and other particles are considered, corresponding $W, T$ identity can also be obtained and the theory is also renormalizable. In fact as we know, as long as theory is gauge invariable, it is certainly renormalizable.

## 2. The elimination of the Higgs mechanism and normalization of Non-Abelian Gauge theory

For simplification, we consider a system composed of gauge field $A_\mu^\alpha$, fermion field $\psi$, and ghost fields $C_\alpha^+$ and $C_\alpha$ at first. Let $S_f$ represent the action of gauge field and fermion field, $S_h$ represent the action of fixed gauge item and $S_g$ represent the action of ghost field. After mass item is added into the actions, the total effective action is $S_{eff} = S_f + S_h + S_g$ with

$$S_f = \int d^4 x \left[ -\overline{\psi} \left( \gamma_\mu \partial_\mu - ig \frac{\tau_\alpha}{2} \gamma_\mu A_\mu^\alpha + m_\psi \right) \psi \right.$$

$$\left. -\frac{1}{4} \left( \partial_\mu A_\nu^\alpha - \partial_\nu A_\mu^\alpha + gf^{\alpha\beta\gamma} A_\mu^\beta A_\nu^\gamma \right)^2 - \frac{1}{2} m_A A_\mu^\alpha A_\mu^\alpha \right] \tag{29}$$

$$S_h = \int d^4 x \left[ -\frac{1}{2\xi} \left( \partial_\mu A_\mu^\alpha \right)^2 \right] \qquad S_g = \int d^4 x C_\alpha^+ R_\mu \left( \delta_{\alpha\beta} \partial_\mu - gf^{\alpha\beta\gamma} A_\mu^\gamma \right) C_\beta \tag{30}$$

Because $A_\mu^\alpha$ is invariable according to this paper, $S_f$ and $S_h$ are invariable under $SU(N)$ gauge transformation. Because $\Delta_F(A_\mu^\alpha)$ is unchanged under gauge transformation, so ghost fields $C_\alpha^+$ and $C_\alpha$ can also be regarded invariable under gauge transformation according to the paper, though they are not in the current theory. In this way, the simplified $B, R, S$ transformations can be written as

$$\delta \overline{\psi} = -i \overline{\psi} \frac{\tau_\alpha}{2} C_\alpha \delta\lambda \qquad\qquad \delta\psi = i \frac{\tau_\alpha}{2} C_\alpha \psi \delta\lambda \tag{31}$$

$$\delta A_\mu^\alpha = 0 \qquad \delta C_\alpha^+ = 0 \qquad \delta C_\alpha = 0 \tag{32}$$

Here $\delta\lambda$ is infinitesimal with $(\delta\lambda)^2 \to 0$. Similarly, we also have $\delta^2 \overline{\psi} = \delta^2 \psi = 0$. The similar generating function of the Green's function, that is unchanged under simplified transformations (31) and

(32), can be written as

$$Z = \int [d\bar{\psi}][d\psi][dA_\mu^\alpha][dC_\alpha^+][dC_\alpha] \exp\{ iS_{eff} +$$

$$+ i\int dx^4 [\bar{\eta}\psi + \bar{\psi}\eta + J_\mu^\alpha A_\mu^\alpha + \zeta_\alpha C_\alpha^+ + \zeta_\alpha^+ C_\alpha + \bar{K}\delta\psi + \delta\bar{\psi}K]\} \tag{33}$$

Here $K$ and $\bar{K}$ are anti-commutative quantities. By considering the facts that integral is independent of variable transformations and $S_{eff}$ is invariable under gauge transformation as well as $\delta^2\bar{\psi} = \delta^2\psi = 0$, we can know that the formula (33) is unchanged under transformations $\bar{\psi} \to \bar{\psi}' = \bar{\psi} + \delta\bar{\psi}$ and $\psi \to \psi' = \psi + \delta\psi$. So we can also write it as

$$Z = \int [d\bar{\psi}][d\psi][dA_\mu^\alpha][dC_\alpha^+][dC_\alpha] \exp\{ iS_{eff} +$$

$$+ i\int dx^4 [\bar{\eta}(\psi + \delta\psi) + (\bar{\psi} + \delta\bar{\psi})\eta + J_\mu^\alpha A_\mu^\alpha + \zeta_\alpha C_\alpha^+ + \zeta_\alpha^+ C_\alpha + \bar{K}\delta\psi + \delta\bar{\psi}K]\} \tag{34}$$

After Eq.(34) minus Eq.(33), we get

$$\int [d\bar{\psi}][d\psi][dA_\mu^\alpha][dC_\alpha^+][dC_\alpha] \int dx^4 [\bar{\eta}\delta\psi + \delta\bar{\psi}\eta]$$

$$\times \exp\{ iS_{eff} + i\int dx^4 (\bar{\eta}\psi + \bar{\psi}\eta + J_\mu^\alpha A_\mu^\alpha + \zeta_\alpha C_\alpha^+ + \zeta_\alpha^+ C_\alpha + \bar{K}\delta\psi + \delta\bar{\psi}K)\} = 0 \tag{35}$$

Let $\delta\psi \to \delta/(i\delta\bar{K})$, $\delta\bar{\psi} \to \delta/(i\delta K)$, the simplified $W$、$T$ identity described by the generating function of the Green's function can be written as

$$\left[\bar{\eta}\frac{\delta}{\delta\bar{K}} + \frac{\delta}{\delta K}\eta\right] Z(\bar{\eta}, \eta, J_\mu^\alpha, \varsigma^+, \varsigma, \bar{K}, K) = 0 \tag{36}$$

The simplified $W,T$ identity described by the generating function of normal vertex angle becomes

$$\frac{\delta\Gamma}{\delta\psi}\frac{\delta\Gamma}{\delta\bar{K}} + \frac{\delta\Gamma}{\delta K}\frac{\delta\Gamma}{\delta\bar{\psi}} = \Gamma * \Gamma = 0 \tag{37}$$

But there is no ghost equation corresponding to $SU(N)$ gauge group. The normalization of single loop approximation is discussed below. It is only a simplified procedure of the current normalization theory. After items $\bar{K}\delta\psi$ and $\delta\bar{\psi}K$ are added into the action, the effective action that is unchanged under simplified $B,R,S$ transformations can be written as

$$S_0 = \int d^4x \left[ -\bar{\psi}\left(\gamma_\mu\partial_\mu - ig\frac{\tau_\alpha}{2}\gamma_\mu A_\mu^\alpha + m_\psi\right)\psi \right.$$

$$-\frac{1}{4}(\partial_\mu A_\nu^\alpha - \partial_\nu A_\mu^\alpha + gf^{\alpha\beta\gamma}A_\mu^\beta A_\nu^\gamma)^2 - \frac{1}{2}m_A^2 A_\mu^\alpha A_\mu^\alpha - \frac{1}{2\xi}(\partial_\mu A_\mu^\alpha)^2$$

$$\left. + C_\alpha^+ R_\mu(\delta_{\alpha\beta}\partial_\mu - gf^{\alpha\beta\gamma}A_\mu^\gamma)C_\beta + ig\bar{K}\frac{\tau_\alpha}{2}C_\alpha\psi\delta\lambda - ig\bar{\psi}\frac{\tau_\alpha}{2}C_\alpha K\delta\lambda \right] \tag{38}$$

Using it to construct the generating function of normal vertex angle, we have $\Gamma[S_0] \approx S_0$ under tree approximation. The process is finite. Therefore, according to Eq.(37), we have $S_0 * S_0 = 0$. For single loop approximation, we can write

$$\Gamma[S_0] = S_0 + \Gamma_1^f[S_0] + \Gamma_1^d[S_0] \tag{39}$$

Here $\Gamma_1^f[S_0]$ is finite but $\Gamma_1^d[S_0]$ is infinite. In order to eliminate infinite, for single loop approximation,

we use $S_0 + \Delta S_0$ to construct the generating function of normal vertex angle

$$\Gamma[S_0 + \Delta S_0] \approx S_0 + \Delta S_0 + \Gamma_1^f[S_0] + \Gamma_1^d[S_0] \qquad (40)$$

By taking $\Delta S_0 = -\Gamma_1^f[S_0]$, the infinite of single loop approximation can be eliminated. It can be proved below that we also have

$$-\Gamma_1^d[S_0] = \sum_\sigma a_\sigma G_\sigma + \hat{S}_0 * F \qquad S_0 * \Gamma_1^d[S_0] = 0 \qquad (41)$$

Here $G_\sigma$ is an invariable quantity of gauge transformation with form

$$\sum_\sigma a_\sigma G_\sigma = \int d^4x \left[ -a_1 \bar{\psi} \left( \gamma_\mu \partial_\mu - ig \frac{\tau_\alpha}{2} \gamma_\mu A_\mu^\alpha \right) \psi - a_2 m_\psi \bar{\psi}\psi \right.$$
$$\left. -\frac{1}{4} a_3 \left( \partial_\mu A_\nu^\alpha - \partial_\nu A_\mu^\alpha + gf^{\alpha\beta\gamma} A_\mu^\beta A_\nu^\gamma \right)^2 - \frac{1}{2} a_4 m_A^2 A_\mu^\alpha A_\mu^\alpha - \frac{1}{2\xi} a_5 \left( \partial_\mu A_\mu^\alpha \right)^2 \right.$$
$$\left. + a_6 C_\alpha^+ R_\mu \left( \delta_{\alpha\beta} \partial_\mu - gf^{\alpha\beta\gamma} A_\mu^\gamma \right) C_\beta \right] \qquad (42)$$

Here $a_i$ are constants containing infinite poles. Because there is no ghost equation, $F$ can be an arbitrary function. We can also write it as similarly

$$F = \int d^4x \left( b_1 \bar{K} \psi + b_2 \bar{\psi} K \right) \qquad (43)$$

Because $G_\sigma$ does not contain $K$ and $\bar{K}$, according to Eq.(33), we have $S_0 \sim \bar{K}\delta\psi + \delta\bar{\psi}K$, so

$$S_0 * G_\sigma = \frac{\delta S_0}{\delta\psi}\frac{\delta G_\sigma}{\delta\bar{K}} + \frac{\delta S_0}{\delta\bar{K}}\frac{\delta G_\sigma}{\delta\psi} + \frac{\delta S_0}{\delta\bar{\psi}}\frac{\delta G_\sigma}{\delta K} + \frac{\delta S_0}{\delta K}\frac{\delta G_\sigma}{\delta\bar{\psi}}$$
$$= \frac{\delta G_\sigma}{\delta\bar{\psi}}\delta\bar{\psi} + \frac{\delta G_\sigma}{\delta\psi}\delta\psi = \delta G_\sigma = 0 \qquad (44)$$

By the anti-commutation relation between $\psi$ and $\bar{\psi}$, it can also be proved as done in the current theory

$$S_0 * (S_0 * F) = 0 \qquad (45)$$

So $\Gamma_1^d[S_0]$ satisfies Eq.(41) and we obtain

$$S_0 + \Delta S_0 = S_0 + \sum_\sigma a_\sigma G_\sigma + \hat{S}_0 * F$$
$$= S_0 + \sum_\sigma a_\sigma G_\sigma + \frac{\delta S_0}{\delta\psi}\frac{\delta F}{\delta\bar{K}} + \frac{\delta S_0}{\delta\bar{K}}\frac{\delta F}{\delta\psi} + \frac{\delta S_0}{\delta\bar{\psi}}\frac{\delta F}{\delta K} + \frac{\delta S_0}{\delta K}\frac{\delta F}{\delta\bar{\psi}} \qquad (46)$$

On the other hand, according to the same method in the current theory, let

$$\bar{\psi}' = \bar{\psi} + \frac{\delta F}{\delta\bar{K}} = (1 + b_1 \bar{\psi}) = Y_{\bar{\psi}} \bar{\psi} \qquad K' = K - \frac{\delta F}{\delta\psi} = (1 - b_1 \bar{K}) = Y_{\bar{\psi}}^{-1} K \qquad (47)$$

$$\psi' = \psi - \frac{\delta F}{\delta K} = (1 - b_2 \psi) = Y_\psi^{-1} \psi \qquad \bar{K}' = \bar{K} + \frac{\delta F}{\delta\bar{\psi}} = (1 + b_2 \bar{K}) = Y_\psi \bar{K} \qquad (48)$$

we can prove [3]

$$S_0(\bar{\psi}', \psi', \bar{K}', K') = S_0(\bar{\psi}, \psi, \bar{K}, K) + \frac{\delta S_0}{\delta\psi}\frac{\delta F}{\delta\bar{K}} + \frac{\delta S_0}{\delta\bar{K}}\frac{\delta F}{\delta\psi} + \frac{\delta S_0}{\delta\bar{\psi}}\frac{\delta F}{\delta K} + \frac{\delta S_0}{\delta K}\frac{\delta F}{\delta\bar{\psi}} \qquad (49)$$

By putting the formula above into Eq.(46), it can be known that the effect of the item $S_0 * F$ is equal to carry out the transformations with the forms of Eqs.(46) and (47) in the action $S_0$. So we can define function $G_\sigma$ by using $\bar{\psi}'$, $\psi'$, $\bar{K}'$, $K'$ at beginning. Then we let $Y_1 = 1 + a_1$, $Y_2 = 1 + a_2$, $Y_3 = 1 + a_3$, $Y_4 = 1 + a_4$, $Y_5 = 1 + a_5$, $Y_6 = 1 + a_6$. In this way, the action of renormalization in single loop process can be written

$$S_1 = S_0 + \Delta S_0 = \int d^4x \left[ -Y_{\bar{\psi}} Y_\psi^{-1} \bar{\psi} \left( \gamma_\mu \partial_\mu - ig \frac{\tau_\alpha}{2} \gamma_\mu A_\mu^\alpha \right) \psi - Y_{\bar{\psi}} Y_\psi^{-1} m_\psi \bar{\psi} \psi \right.$$

$$- \frac{1}{4} \left( \partial_\mu A_\nu^\alpha - \partial_\nu A_\mu^\alpha + g f^{\alpha\beta\gamma} A_\mu^\beta A_\nu^\gamma \right)^2 - \frac{1}{2} m_A^2 A_\mu^\alpha A_\mu^\alpha - \frac{1}{2\xi} \left( \partial_\mu A_\mu^\alpha \right)^2$$

$$+ C_\alpha^+ R_\mu \partial_\mu C_\alpha - g f^{\alpha\beta\gamma} C_\alpha^+ R_\mu A_\mu^\beta C_\gamma + ig \bar{K} \frac{\tau_\alpha}{2} C_\alpha \psi \delta\lambda - ig \bar{\psi} \frac{\tau_\alpha}{2} C_\alpha K \delta\lambda$$

$$- (Y_1 - 1) Y_{\bar{\psi}} Y_\psi^{-1} \bar{\psi} \left( \gamma_\mu \partial_\mu - ig \frac{\tau_\alpha}{2} \gamma_\mu A_\mu^\alpha \right) \psi - (Y_2 - 1) Y_{\bar{\psi}} Y_\psi^{-1} m_\psi \bar{\psi} \psi$$

$$- \frac{1}{4} (Y_3 - 1) \left( \partial_\mu A_\nu^\alpha - \partial_\nu A_\mu^\alpha + g f^{\alpha\beta\gamma} A_\mu^\beta A_\nu^\gamma \right)^2 - \frac{1}{2} (Y_4 - 1) m_A^2 A_\mu^\alpha A_\mu^\alpha$$

$$\left. - \frac{1}{2\xi} (Y_5 - 1) \left( \partial_\mu A_\mu^\alpha \right)^2 + (Y_6 - 1) \left( C_\alpha^+ R_\mu \partial_\mu C_\alpha - g f^{\alpha\beta\gamma} C_\alpha^+ R_\mu A_\mu^\beta C_\gamma \right) \right]$$

$$= \int d^4x \left[ -Y_1 Y_{\bar{\psi}} Y_\psi^{-1} \bar{\psi} \left( \gamma_\mu \partial_\mu - ig \frac{\tau_\alpha}{2} \gamma_\mu A_\mu^\alpha \right) \psi - Y_2 Y_{\bar{\psi}} Y_\psi^{-1} m_\psi \bar{\psi} \psi \right.$$

$$- \frac{1}{4} Y_3 \left( \partial_\mu A_\nu^\alpha - \partial_\nu A_\mu^\alpha \right)^2 - \frac{1}{2} Y_3 g \left( \partial_\mu A_\nu^\alpha - \partial_\nu A_\mu^\alpha \right) f^{\alpha\beta\gamma} A_\mu^\beta A_\nu^\gamma$$

$$- \frac{1}{4} Y_3 \left( g f^{\alpha\beta\gamma} A_\mu^\beta A_\nu^\gamma \right)^2 - \frac{1}{2} Y_4 m_A^2 A_\mu^\alpha A_\mu^\alpha - \frac{1}{2\xi} Y_5 \left( \partial_\mu A_\mu^\alpha \right)^2$$

$$\left. + Y_6 \left( C_\alpha^+ R_\mu \partial_\mu C_\alpha - g f^{\alpha\beta\gamma} C_\alpha^+ R_\mu A_\mu^\beta C_\gamma \right) + ig \bar{K} \frac{\tau_\alpha}{2} C_\alpha \psi \delta\lambda - ig \bar{\psi} \frac{\tau_\alpha}{2} C_\alpha K \delta\lambda \right] \quad (50)$$

On the other hand, when the action is described by the nude quantities, we have

$$S_1 = \int d^4x \left[ -\bar{\psi}_0 \left( \gamma_\mu \partial_\mu - ig_0 \frac{\tau_\alpha}{2} \gamma_\mu A_{0\mu}^\alpha + m_{0\psi} \right) \psi_0 \right.$$

$$- \frac{1}{4} \left( \partial_\mu A_{0\nu}^\alpha - \partial_\nu A_{0\mu}^\alpha + g_0 f^{\alpha\beta\gamma} A_{0\mu}^\beta A_{0\nu}^\gamma \right)^2 - \frac{1}{2} m_{0A}^2 A_{0\mu}^\alpha A_{0\mu}^\alpha - \frac{1}{2\xi_0} \left( \partial_\mu A_{0\mu}^\alpha \right)^2$$

$$\left. + C_{0\alpha}^+ R_\mu \partial_\mu C_{0\alpha} - g_0 f^{\alpha\beta\gamma} C_{0\alpha}^+ R_\mu A_{0\mu}^\beta C_{0\gamma} + ig_0 \bar{K}_0 \frac{\tau_\alpha}{2} C_{0\alpha} \psi_0 \delta\lambda - ig_0 \bar{\psi}_0 \frac{\tau_\alpha}{2} C_{0\alpha} K_0 \delta\lambda \right] \quad (51)$$

Let $\psi_0 = \sqrt{Z_2} \psi$, $\bar{\psi}_0 = \sqrt{Z_2} \bar{\psi}$, $A_{0\mu}^\alpha = \sqrt{Z_3} A_\mu^\alpha$, $C_{0\alpha}^+ = \sqrt{\tilde{Z}_3} C_\alpha^+$, $C_{0\alpha} = \sqrt{\tilde{Z}_3} C_\alpha$, $\bar{K}_0 = \sqrt{Z_K} \bar{K}$, $K_0 = \sqrt{Z_K} K$, $g_0 = Z_{gi} g$, $m_{0\psi} = Z_{m\psi} m_\psi$, $m_{0A} = Z_{mA} m_A$, $\xi_0 = Z_\xi \xi$, the formula above becomes

$$S_1 = \int d^4x \left[ -Z_2 \bar{\psi} \left( \gamma_\mu \partial_\mu - i Z_{g1} \sqrt{Z_3} g \frac{\tau_\alpha}{2} \gamma_\mu A_\mu^\alpha + Z_{m\psi} m_\psi \right) \psi \right.$$

$$- \frac{1}{4} Z_3 \left( \partial_\mu A_\nu^\alpha - \partial_\nu A_\mu^\alpha \right)^2 - \frac{1}{2} Z_{g2} Z_3^{3/2} g \left( \partial_\mu A_\nu^\alpha - \partial_\nu A_\mu^\alpha \right) f^{\alpha\beta\gamma} A_\mu^\beta A_\nu^\gamma$$

$$- \frac{1}{4} Z_{g3}^2 Z_3^2 \left( g f^{\alpha\beta\gamma} A_\mu^\beta A_\nu^\gamma \right)^2 - \frac{1}{2} Z_{mA}^2 Z_3 m_A^2 A_\mu^\alpha A_\mu^\alpha - \frac{1}{2\xi} Z_\xi Z_3 \left( \partial_\mu A_\mu^\alpha \right)^2$$

$$+ \tilde{Z}_3 R_\mu C_\alpha^+ \partial_\mu C_\alpha - Z_{g4} \tilde{Z}_3 \sqrt{Z_3} g f^{\alpha\beta\gamma} C_\alpha^+ R_\mu A_\mu^\beta C_\gamma$$

$$+ i Z_{g5} Z_K \sqrt{\tilde{Z}_3 Z_2} g \overline{K} \frac{\tau_\alpha}{2} C_\alpha \psi \delta\lambda - i Z_{g6} Z_K \sqrt{\tilde{Z}_3 Z_2} g \overline{\psi} \frac{\tau_\alpha}{2} C_\alpha K \delta\lambda \Big] \tag{52}$$

Comparing the corresponding items in Eqs.(51) and (52), we get

$$Z_2 = Y_1 Y_{\overline{\psi}} Y_\psi^{-1} \qquad Z_{g1} \sqrt{Z_3} = 1 \qquad Z_2 Z_{m\psi} = Y_2 Y_{\overline{\psi}} Y_\psi^{-1} \qquad Z_3 = Y_3$$

$$Z_{g2} \sqrt{Z_3} = 1 \qquad Z_{g3}^2 Z_3 = 1 \qquad Z_{mA}^2 Z_3 = Y_4 \qquad Z_\xi Z_3 = Y_5$$

$$\tilde{Z}_3 = Y_6 \qquad Z_{g4} \sqrt{Z_3} = 1 \qquad Z_{g5} Z_K \sqrt{\tilde{Z}_3 Z_2} = 1 \qquad Z_{g6} = Z_{g5} \tag{53}$$

It can be obtained immediately

$$Z_{g1} = Z_{g2} = Z_{g3} = Z_{g4} = Z_g = \frac{1}{\sqrt{Z_3}} = \frac{1}{\sqrt{Y_3}} \tag{54}$$

By taking $Z_K = Y_3 / \sqrt{\tilde{Z}_3 Z_2}$ (similar to the current theory), we have $Z_{g5} = Z_{g6} = Z_g$. Therefore, the renormalization interaction constants in all items are the same so that renormalization is possible. So for the process of single loop approximation, according to the paper, renormalization constants are taken as

$$Z_2 = Y_1 Y_{\overline{\psi}} Y_\psi^{-1} \qquad Z_3 = Y_3 \qquad \tilde{Z}_3 = Y_6 \qquad Z_g = \frac{1}{\sqrt{Y_3}}$$

$$Z_{m\psi} = \frac{Y_2}{Y_1} \qquad Z_{mA} = \sqrt{\frac{Y_4}{Y_3}} \qquad Z_\xi = \frac{Y_5}{Y_3} \qquad Z_K = \frac{\sqrt{Y_3 Y_\psi}}{\sqrt{Y_6 Y_1 Y_{\overline{\psi}}}} \tag{55}$$

In fact, because there is no the restriction of ghost equation for $SU(N)$ fields according to the paper, the function $F$ in Eq.(41) can be arbitrary. For simplification, we can take $F = 0$ directly so that it is unnecessary for us to introduce Eq.(47) and (48) again. In this case we have $Y_{\overline{\psi}} = Y_\psi^{-1} = 1$ in Eq.(53) and (55). For higher order processes, renormalization can also be carried out by the similar procedure in the current theory. But it is unnecessary for us to discuss any more here.

## 3. The elimination of the Higgs mechanism and normalization of weak-electric united theory

The mass item's gauge transformation in the united theory of weak-electric interaction is discussed at last. We only discuss the transformation of lepton field's mass items. The result is suitable to quark fields. In the united theory, we use chiral fields to describe weak interaction. The transformation rules of left hand and right hand fields under $SU(2) \times U(1)$ gauge transformation are

$$L \to L' = exp\left(-i \frac{\overline{\theta} \cdot \overline{\tau}}{2} + i \frac{\theta}{2}\right) L \qquad L = \begin{vmatrix} v_L \\ l_L \end{vmatrix} \tag{56}$$

$$l_R \to exp(i\theta) l_R \qquad l_L = \frac{1}{2}(1 + \gamma_5) l \qquad l_R = \frac{1}{2}(1 - \gamma_5) l \tag{57}$$

The Lagrangian of free lepton field without mass item is

$$\mathcal{L}_{l0} = -\bar{L}\gamma_\mu \partial_\mu L - \bar{l}_R \gamma_\mu \partial_\mu l_R \tag{58}$$

Because the transformation rule of left hand field is different from that of right hand field, the mass item of lepton field with form

$$m_l \bar{l} l = m_l (\bar{l}_L l_R + \bar{l}_R l_L) \tag{59}$$

can't not keep unchanged under $SU(2) \times U(1)$ transformation. So the mass items of lepton fields can't yet be added into the Lagrangian directly according to the current theory. The Higgs mechanics is needed. It is proved below that by introducing some proper restriction relations between group parameters, the Higgs mechanics is also unnecessary.

According to Eq.(56), we have infinitesimal transformations

$$v'_L \approx \left[1 - \frac{i}{2}(\theta_3 - \theta)\right] v_L - \frac{i}{2}(\theta_1 - i\theta_2) l_L \qquad v' \approx \left[1 - \frac{i}{2}(\theta_3 - \theta)\right] v - \frac{i}{2}(\theta_1 - i\theta_2) l \tag{60}$$

$$l'_L \approx -\frac{i}{2}(\theta_1 + i\theta_2) v_L + \left[1 - \frac{i}{2}(\theta_3 + \theta)\right] l_L \qquad l' \approx -\frac{i}{2}(\theta_1 + i\theta_2) v + \left[1 - \frac{i}{2}(\theta_3 + \theta)\right] l \tag{61}$$

If choosing restriction relation $\theta_1 = -i\theta_2$, we get

$$l' \approx \left[1 - \frac{i}{2}(\theta_3 + \theta)\right] l \approx \exp\left[-\frac{i}{2}(\theta_3 + \theta)\right] l \tag{62}$$

In this case, we have $\bar{l}' l' = \bar{l} l$, the lepton mass item can keep unchanged under $SU(2) \times U(1)$ transformation and can be added into the Lagrangian directly.

For the transformation of gauge field's mass items, the relations between mass eigen states and non-mass eigen states of gauge particles are

$$W^+_\mu = \frac{1}{\sqrt{2}}(A^1_\mu + iA^2_\mu) \qquad W^-_\mu = \frac{1}{\sqrt{2}}(A^1_\mu - iA^2_\mu) \tag{63}$$

$$Z_\mu = \cos\vartheta_w A^3_\mu - \sin\vartheta_w B_\mu \qquad A_\mu = \sin\vartheta_w A^3_\mu + \cos\vartheta_w B_\mu \tag{64}$$

Here $\vartheta_w$ is the Weinberg angle, $A_\mu$ is electromagnetic field, $B_\mu$ field has no mass. We can get

$$\frac{1}{2} m_A^2 A^\alpha_\mu A^\alpha_\mu = m_A^2 W^+_\mu W^-_\mu + \frac{1}{2} m_A^2 \cos\vartheta_w^2 Z_\mu Z_\mu + Q \tag{65}$$

$$Q = \frac{1}{2} m_A^2 \sin^2\vartheta_w (A_\mu)^2 + m_A^2 \sin\vartheta_w \cos\vartheta_w A_\mu Z_\mu \tag{66}$$

In the formula, $m_A \sin\vartheta_w$ is photon's mass and product item $A_\mu Z_\mu$ represent two point's interaction. Because theses two items do not exist actually, we should cancel them in the action. Just as taking $R_\xi$ gauge in the current theory, we take following gauge

$$F^\alpha(A^\alpha_\mu) = \partial_\mu A^\alpha_\mu + R^\alpha \qquad R^\alpha = -\partial_\mu A^\alpha_\mu \pm \sqrt{1 - 2\xi_A Q / \partial_\mu A^\alpha_\mu} \tag{67}$$

So the gauge fixed item can be written as

$$S_h = \int d^4 x \left[-\frac{1}{2\xi_A}(\partial_\mu A^\alpha_\mu)^2 + Q\right] \tag{68}$$

In this way, the superfluous factor $Q$ appearing in the action can be canceled. Thus, let $m_A = m_w$, we

have

$$\frac{1}{2}m_w^2 A_\mu^\alpha A_\mu^\alpha \sim m_w^2 W_\mu^+ W_\mu^- + \frac{1}{2} m_w^2 \cos\vartheta_w^2 Z_\mu Z_\mu \tag{69}$$

Because $m_w \cos\theta_w$ is $Z^0$ particle's mass actually, we have

$$m_Z = m_w \cos\vartheta_w \tag{70}$$

By calculating the low order process of $\mu^-$ decay and comparing the result with the Fermi theory, we can also get $G/\sqrt{2} = g^2/(8m_w^2)$, from which we can decide the masses of $W^\pm$ particles. Then from Eq.(70), $Z^0$ particle's mass can also be determined. The result is completely the same as that in the current theory in which the Higgs mechanics is used. When mass eigen states are used, the gauge transformation of mass items is

$$m_w^2 W_\mu'^+ W_\mu'^- + \frac{1}{2} m_Z^2 Z_\mu' Z_\mu' = m_w^2 W_\mu^+ W_\mu^- + \frac{1}{2} m_Z^2 Z_\mu Z_\mu$$

$$+ \frac{\sin\vartheta_w}{2g}\left(2\cos\vartheta_w A_\mu^3 - 2\sin\vartheta_w B_\mu + \frac{\sin\vartheta_w}{g}\partial_\mu\theta\right)\partial_\mu\theta \tag{71}$$

It can't keep unchanged under gauge transformation. In order to let it unchanged with

$$m_w^2 W_\mu'^+ W_\mu'^- + \frac{1}{2} m_Z^2 Z_\mu' Z_\mu' = m_w^2 W_\mu^+ W_\mu^- + \frac{1}{2} m_Z^2 Z_\mu Z_\mu \tag{72}$$

we can take

$$\partial_\mu \theta = -\frac{2g}{\sin\vartheta_w}\left(\cos\vartheta_w A_\mu^3 - \sin\vartheta_w B_\mu\right) = -\frac{2g}{\sin\vartheta_w} Z_\mu \tag{73}$$

So in order to keep the mass items represented by mass eigen states unchanged under $SU(2) \times U(1)$ transformation, the form of group parameter $\theta$ can not yet be arbitrary. Eq.(73) should be satisfied. It is noted that according to the definition in Eq.(9), group parameter $\theta$ is finite. For infinitesimal transformation, we should let $\theta \to \theta\delta\lambda$ with $\partial_\mu\theta\delta\lambda = -2gZ_\mu\delta\lambda/\sin\vartheta_w$. In this way, the mass items of particles $W^\pm$ and $Z^0$ can be added into the Lagrangian directly without violating $SU(2) \times U(1)$ gauge invariability.

On the other hand, because it is unnecessary for us to introduce ghost field corresponding to $U(1)$ gauge field, there is no the gauge fixed item corresponding to $B_\mu$. Thus, when non-mass eigen states are used, we have the gauge invariable action of electro-weak united theory

$$S_0 = \int d^4x \left[ -\bar{L}\left(\gamma_\mu\partial_\mu - ig\frac{\tau^\alpha}{2}\gamma_\mu A_\mu^\alpha + ig'\frac{1}{2}\gamma_\mu B_\mu\right)L - \bar{l}_R(\gamma_\mu\partial_\mu + ig'\gamma_\mu B_\mu)l_R \right.$$

$$-\frac{1}{4}\left(\partial_\mu A_\nu^\alpha - \partial_\nu A_\mu^\alpha + gf^{\alpha\beta\gamma}A_\mu^\beta A_\nu^\gamma\right)^2 - \frac{1}{4}\left(\partial_\mu B_\nu - \partial_\nu B_\mu\right)^2$$

$$-\frac{1}{2}m_A^2 A_\mu^\alpha A_\mu^\alpha - m_l\left(\bar{l}_L l_R + \bar{l}_R l_L\right) - \frac{1}{2\xi_A}\left(\partial_\mu A_\nu^\alpha\right)^2 - \frac{1}{2\xi_B}\left(\partial_\mu B_\mu\right)^2$$

$$+ C_\alpha^+ R_\mu \partial_\mu C_\alpha - gf^{\alpha\beta\gamma}C_\alpha^+ R_\mu A_\mu^\beta C_\gamma + C^+\partial^2 C + \bar{K}_1\delta v_L + \delta\bar{v}_L K_1$$

$$\left. + \bar{K}_2\delta l_L + \delta\bar{l}_L K_2 + \bar{K}_3\delta l_R + \delta\bar{l}_R K_3 + \bar{K}_4\delta B_\mu + \delta\bar{B}_\mu K_4 \right] \tag{74}$$

According to Eqs. (9), (60) and (61), the infinitesimal transformations are

$$\delta \bar{v}_L = \frac{i}{2}(\theta_3 - \theta)\bar{v}_L + i\theta_1 \bar{l}_L \qquad \delta \bar{l}_L = \frac{i}{2}(\theta_3 + \theta)\bar{l}_L \qquad \delta \bar{l}_R = i\theta \bar{l}_R$$

$$\delta v_L = -\frac{i}{2}(\theta_3 - \theta)v_L - i\theta_1 l_L \qquad \delta l_L = -\frac{i}{2}(\theta_3 + \theta)l_L \qquad \delta l_R = i\theta l_R$$

$$\delta A_\mu^\alpha = 0 \qquad \delta B_\mu = -\frac{1}{g}\partial_\mu \theta \qquad \delta C_\alpha^+ = \delta C_\alpha = \delta C = 0 \qquad \delta C^+ = \xi_B \partial_\mu B_\mu \qquad (75)$$

Let $\theta_i = C_i \delta\lambda$, $\theta = C\delta\lambda$ similarly, we have $(\delta\lambda)^2 \to 0$ so that $\delta^2 v_L = \delta^2 l_L = \delta^2 B_\mu \to 0$. Put them into Eq.(74), by the same method shown before, renormalization calculation can be done. If mass eigen states are used, by the transformation relations Eqs.(63) and (64), we can also get the action which is also invariable under $SU(2) \times U(1)$ transformation

$$S_0 = \int d^4x \left\{ \mathcal{L}_0(W_\mu^\pm, Z_\mu, A_\mu, l, v) + i\frac{g}{\sqrt{2}}\left[W_\mu^+ \bar{v} \gamma_\mu (1+\gamma_5) l + W_\mu^- \bar{l} \gamma_\mu (1+\gamma_5) v\right] \right.$$

$$-i\frac{\sqrt{g^2+g'^2}}{4} Z_\mu \left[\bar{v}\gamma_\mu(1+\gamma_5)v + \bar{l}\gamma_\mu(4\sin\theta_w - 1 - \gamma_5) l\right] - i\frac{gg'}{\sqrt{g^2+g'^2}} A_\mu \bar{l}\gamma_\mu l$$

$$- m_l \bar{l}l - m_w^2 W^+W^- - \frac{1}{2}m_Z^2 Z_\mu Z_\mu - \frac{1}{2\xi_A}\left[\partial_\mu A_\mu^\alpha(W_\mu^\pm, Z_\mu, A_\mu)\right]^2$$

$$- \frac{1}{2\xi_B}\left[\partial_\mu B_\mu(W_\mu^\pm, Z_\mu, A_\mu)\right]^2 + C_\alpha^+ R_\mu \partial_\mu C_\alpha - gf^{\alpha\beta\gamma} C_\alpha^+ R_\mu A_\mu^\beta(W_\mu^\pm, Z_\mu, A_\mu) C_\gamma + C_\alpha^+ \partial^2 C_\alpha$$

$$\left. + \bar{K}_1 \delta v + \delta \bar{v} K_1 + \bar{K}_2 \delta l_L + \delta \bar{l}_L K_2 + \bar{K}_3 \delta l_R + \delta \bar{l}_R K_3 + \bar{K}_4 \delta B_\mu + \delta \bar{B}_\mu K_4 \right\} \qquad (76)$$

In the formula, $\mathcal{L}_0$ is the Lagrangian of free fields without containing mass items. In this case, the transformation rules of various fields becomes

$$\delta v = -\frac{i}{2}(\theta_3 - \theta)v - i\theta_1 l \qquad \delta \bar{v} = \frac{i}{2}(\theta_3 - \theta)\bar{v} + i\theta_1 \bar{l} \qquad \delta l = -\frac{i}{2}(\theta_3 + \theta)l$$

$$\delta \bar{l} = \frac{i}{2}(\theta_3 + \theta)\bar{l} \qquad \delta W_\mu^+ = 0 \qquad \delta W_\mu^- = 0 \qquad \delta Z_\mu = -\sin\theta_w \delta B_\mu$$

$$\delta A_\mu = \frac{1}{g}\cos\theta_w \delta B_\mu \qquad \delta C_\alpha^+ = \delta C_\alpha = \delta C = 0 \qquad \delta C^+ = \xi_B \partial_\mu B_\mu \qquad (77)$$

Let $\theta_i = C_i \delta\lambda$, $\theta = C\delta\lambda$ similarly, we have $(\delta\lambda)^2 \to 0$, $\delta^2 v = \delta^2 l = \delta^2 Z_\mu = \delta^2 A_\mu \to 0$. The same renormalization calculation can be carried out. That is to say, when mass eigen states are used to construct the effect action of electro-weak united theory, the theory is still gauge invariable and renormalizable. However, this is impossible in the current theory to use the Higgs mechanics. After the spontaneous breaking of vacuum symmetry is completed and gauge particles obtain masses, the effective actions would have no gauge symmetries again.

**Discussion** In order to make the theory of non-Abelian gauge field consistent and rational, we should consider the invariability of motion equation besides the Lagrangian. The introduction of the restriction condition Eq.(18) is an inevitable result that the gauge transformations of non-Abelian gauge field's motion equations must satisfy. We should give up the principle of completely local gauge invariability and adopt the principle of incompletely local gauge invariability. The result is that we can make the theory of non-Abelian gauge field self-consistent without introducing the Higgs mechanism. In this way, the Higgs mechanics becomes surplus. Because the Higgs particles can't be found up to now, it is still a big problem whether or not they exist. At present, some theories have been put forward to replace the Higgs particles. For example, the Higgs particles are regarded as the bounding states of some new positive and anti-quark particles. But all these theories have some difficult problems. It can be said that the scheme provided in this paper is simplest method to solve this problem without increasing any new particles or hypotheses. The introduction of the restriction condition does not cause any inconsistency that contradicts with current experiments. The revision only means to cancel all content relative to the Higgs particles and remain others in the standard model of particle physics. The description of gauge field theory can also become more symmetrical and simple.

The result above can be used to solve the so-called $CP$ violation of strong interaction. Speaking simply, according to the formula (4), under the condition of pure gauge $A_\mu^\alpha(x)\big|_{r\to\infty}=0$, we have

$$A_\mu'^\alpha(x) = U(x)A_\mu^\alpha(x)U^{-1}(x) + \frac{i}{g}T^\alpha U(x)\partial_\mu U(x) \to \frac{i}{g}T^\alpha U(x)\partial_\mu U(x) \tag{78}$$

Based on the formula above, the so-called instanton solution and the problem of $\theta$ vacuum[1] are caused. The effect of $\theta$ vacuum is equal to introduce an additive item into the Lagrange of strong interaction with the invariability of gauge transformation. But this item would cause big $CP$ violation and big electric dipole moment for neutron which are considered not to exist in experiments.

However, according to the discussion above, because the gauge potential itself is unchanged under gauge transformation, we always have $A_\mu'^\alpha(x) = A_\mu^\alpha(x)$ in any condition. Under the pure gauge condition when $A_\mu^\alpha(x)\big|_{r\to\infty}=0$, we still have $A_\mu'^\alpha(x)=0$ so that relation $A_\mu'^\alpha(x) = iT^\alpha U(x)\partial_\mu U(x)/g$ does not exist actually. In this way, the so-called instanton solution and the problem of $\theta$ vacuum, as well as the so-called strong $CP$ violation do not exist too. We also do not need the hypothesis of axion again.